\documentclass[twocolumn,preprintnumbers,showpacs,showkeys,amsmath,amssymb]{revtex4}
\usepackage{graphicx}
\usepackage{dcolumn}
\usepackage{bm}
\begin{document}
\title{Equivalence of a one-dimensional driven-diffusive system and an equilibrium two-dimensional walk model}
\author{Farhad H. Jafarpour}
\email{farhad@ipm.ir}
\author{Somayeh Zeraati}
\affiliation{Physics Department, Bu-Ali Sina University, 65174-4161 Hamedan, Iran}
\date{\today}
\begin{abstract}
It is known that a single product shock measure in some of one-dimensional driven-diffusive systems with nearest-neighbor interactions might evolve in time quite similar to a random walker moving on a one-dimensional lattice with reflecting boundaries. The non-equilibrium steady-state of the system in this case can be written in terms of a linear superposition of such uncorrelated shocks. Equivalently, one can write the steady-state of this system using a matrix-product approach with two-dimensional matrices. In this paper we introduce an equilibrium two-dimensional one-transit walk model and find its partition function using a transfer matrix method. We will show that there is a direct connection between the partition functions of these two systems. We will explicitly show that in the steady-state the transfer matrix of the one-transit walk model is related to the matrix representation of the algebra of the driven-diffusive model through a similarity transformation. The physical quantities are also related through the same transformation.
\end{abstract}
\maketitle
\section{Introduction}
During the last couple of years, the time evolution of a single product shock measure in one-dimensional driven-diffusive systems \cite{Schm95,Priv97,Schu00} has been studied in a series of papers \cite{BS02,KJS03,PS04,RS04,TS061,TS062}. It has been observed that by applying some constraints on the microscopic reaction rates of the system the shock front might have simple random walk dynamics while reflecting from the boundaries of the lattice. In this case the steady-state of the system can be easily obtained by considering a linear superposition of such product measures. Three families of two-state systems with nearest-neighbor interactions have been introduced and studied in \cite{KJS03}. This has been also generalized to the multi-species systems with long range interactions and even the systems with discrete time updating scheme \cite{JM09}. \\
The study of the steady-state properties of one-dimensional driven-diffusive systems has a long history. Different approaches have been used in order to find the stationary-state probability distribution function of these systems ranging from the Bethe ansatz to the matrix-product approach. In the latter scenario one writes the stationary-state of the system as a matrix element of product of non-commuting operators associated with different states of each lattice site of the system. The algebraic relation between these operators, sometimes called the algebra of the system, might have finite- or infinite-dimensional matrix representations \cite{BE07}. In an attempt to understand the nature of the matrix representations of these algebras, it has been observed that for the systems in which a single product shock measure has a simple random walk dynamics under the time evolution generated by the so-called Hamiltonian of the system, the stationary-state probability distribution function can be expressed by a two-dimensional matrix representation. These matrices contain all of the information about the hopping rates of the shock front and also the densities of the particles on the left and the right hand side of the shock position. It has been confirmed that the conditions under which the product shock measure has a simple random walk dynamics are exactly those for the existence of a two-dimensional matrix representation for the algebra of the system \cite{JM07}. \\
Following the paper by Arndt \cite{A00} investigations have shown that some of the concepts used in the equilibrium statistical mechanics can be extended to the out-of-equilibrium systems. These concepts include the Yang-Lee description of phase transitions in equilibrium systems. It is known that by defining an {\it ad hoc} partition function in terms of the microscopic reaction rates of an out-of-equilibrium system, one can apply the Yang-Lee approach to spot the transition points. On the other hand, in \cite{BE04,BGR04,BJJK04s} the connection between the steady-state of a one-dimensional driven-diffusive system i.e. the Asymmetric Simple Exclusion Process (ASEP) in random sequential updating scheme with that of a one-transit walk model was investigated. The investigations have been also extended to the case of ASEP with parallel updating scheme in \cite{BJJK04p}.\\
The authors in \cite{BGR04} have shown that by introducing two fugacities associated with the densities of contact points in the equilibrium one-transit walk model one can clearly explain the phases in the walk model. On the other hand, they have shown that if one replaces the two fugacities with the boundary rates of ASEP, a better understanding of  the phase diagram of this non-equilibrium system can be obtained. In the present paper we extend the above mentioned ideas to the case where the shocks with simple random walk dynamics appear in the most general one-dimensional driven-diffusive system. We define a one-dimensional driven-diffusive lattice model with open boundaries in which it is assumed that a single product shock measure can move on the lattice with random walk dynamics. One can introduce an {\it ad hoc} partition function for this model in terms of its microscopic reaction rates. We also introduce an equilibrium one-transit walk model. As in \cite{BGR04} the partition function of this model can be calculated by defining proper fugacities. We aim to show that the physical quantities in both models are closely related. We believe that the connection between these two systems in the steady-state helps us understand the nature of applicability of equilibrium concepts to the non-equilibrium systems.\\
This paper is organized as follows: We start with the definition of the walk model and calculate some of its relevant physical quantities in terms of two proper fugacities. In order to find the partition function of this model we will use the transfer matrix method. Then we consider a general one-dimensional driven-diffusive system and show how its steady-state probability distribution function can be calculated using a matrix-product method and a two-dimensional matrix representation. We finally conclude by showing that the physical quantities of these two systems are closely related if one relates the hopping rates of the shock front in the driven-diffusive system with the fugacities in the walk model.
\section{The walk model}
\begin{figure}[htbp]
\begin{center}
\begin{picture}(140,100)
\put(-30,50){\line(1,0){200}}
\put(-30,50){\line(1,1){20}}
\put(-10,70){\line(1,-1){20}}
\put(10,50){\line(1,1){20}}
\put(30,70){\line(1,-1){20}}
\put(50,50){\line(1,1){20}}
\put(70,70){\line(1,-1){40}}
\put(110,30){\line(1,1){20}}
\put(130,50){\line(1,-1){20}}
\put(150,30){\line(1,1){20}}
\put(-40,40){\footnotesize $(0,0)$}
\put(-20,75){\footnotesize $(1,1)$}
\put(0,40){\footnotesize $(2,0)$}
\put(130,15){\footnotesize $(2n-1,-1)$}
\put(155,55){\footnotesize $(2n,0)$}
\end{picture}
\caption[fig1]{A typical one-transit walk.}
\label{fig1}
\end{center}
\end{figure}
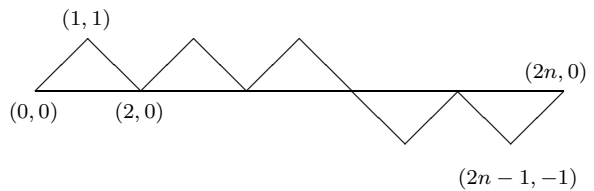

\begin{figure}[htbp]
\begin{center}
\begin{picture}(140,100)
\put(0,50){\line(1,0){120}}
\put(0,50){\line(1,1){30}}
\put(30,80){\line(1,-1){60}}
\put(90,20){\line(1,1){30}}
\put(0,60){\footnotesize $\downarrow$}
\put(-5,70){\footnotesize $\vert 0^+ \rangle$}
\put(30,85){\footnotesize $\downarrow$}
\put(25,95){\footnotesize $\vert 1^+ \rangle$}
\put(60,60){\footnotesize $\downarrow$}
\put(55,70){\footnotesize $\vert 0^+ \rangle$}
\put(60,40){\footnotesize $\uparrow$}
\put(55,30){\footnotesize $\vert 0^- \rangle$}
\put(90,10){\footnotesize $\uparrow$}
\put(85,0){\footnotesize $\vert 1^- \rangle$}
\put(120,40){\footnotesize $\uparrow$}
\put(115,30){\footnotesize $\vert 0^- \rangle$}
\end{picture}
\caption[fig2]{Assigning the base vectors to the vertices}
\label{fig2}
\end{center}
\end{figure}
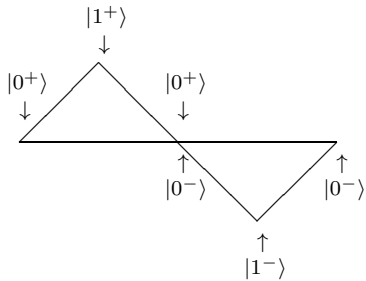

Consider a random walker which moves on a two-dimensional path. Each path is defined on a diagonally-rotated square lattice. The paths start at $(0,0)$ and end at $(2n,0)$. The walker can only move in the north-east (NE) or in the south-east (SE) direction. The walker can never take two consecutive steps to NE while it might take two consequent steps to SE only once. Each path either crosses the horizontal line only once (in this case the path can be factorized into two Dyck paths) or never crosses it (in this case the path is a single Dyck path). According to our definition there are only $n+1$ different paths of this type. A typical path is illustrated in FIG. (\ref{fig1}). As can be seen, the height of each step can be $+1$, $0$ or $-1$. A fugacity $z_1$ is given to each down step and $z_2$ to each up step, except those ending on the $x$-axis. The partition function of this model which contains $2n$ steps can be easily written as
\begin{equation}
\label{PF}
Z_{n}(z_1,z_2)=(z_1z_2)^n\widetilde{Z}_{n}(z_1,z_2)
\end{equation}
in which
\begin{equation}
\label{RPF}
\widetilde{Z}_{n}(z_1,z_2)=\sum_{q=0}^{n}z_1^{-q}z_2^{-n+q}=\frac{z_1^{-n}z_2-z_2^{-n}z_1}{z_2-z_1}.
\end{equation}
The expression (\ref{RPF}) has a simple interpretation given in \cite{BJJK04s}. The weight $z_1^{-q}z_2^{-n+q}$, associated with a path containing $q$ contacts with the $x$-axis from above, is obtained by assigning a factor $z_1^{-1}$ to each contact with the $x$-axis from above, excluding an initial upward step, and a factor $z_2^{-1}$ to each contact from below, excluding a final upward step. As explained in \cite{BJJK04s} the prefactor $(z_1z_2)^n$ does not change the critical behavior of the walk model.\\
In what follows we will show that the partition function $\widetilde{Z}_{n}(z_1,z_2)$ can be written as
\begin{equation}
\label{RPFTM}
\widetilde{Z}_{n}(z_1,z_2)=\langle L \vert T^n \vert R \rangle
\end{equation}
in which $T$ is a two-step transfer matrix. We associate $T^o$ to each odd step and $T^e$ to each even step and write $T=T^oT^e$. In order to find the transfer matrix $T$ and the vectors $\vert R \rangle$ and $\langle L \vert$ we first introduce four base vectors $\vert s_1 \rangle=\vert 1^+ \rangle$, $\vert s_2 \rangle = \vert 1^- \rangle$, $\vert s'_1 \rangle=\vert 0^+ \rangle$ and $\vert s'_2 \rangle=\vert 0^- \rangle$ with the following properties
\begin{equation}
\sum_{\{s_i\}}\vert s_i\rangle \langle s_i \vert =
\sum_{\{s'_i\}}\vert s'_i\rangle \langle s'_i \vert =\mathcal{I}
\end{equation}
in which $\mathcal{I}$ is an identity $2 \times 2$ matrix. Using the completeness of the base vectors the partition function $\widetilde{Z}_{n}(z_1,z_2)$ can be expand as follows
\begin{widetext}
\begin{equation}
\widetilde{Z}_{n}(z_1,z_2)=\sum_{\{s'_i,s'_j,\cdots,s'_{n+1},s_i,s_j,\cdots,s_{n}\}}\langle L \vert s'_i \rangle \langle s'_i \vert T^o \vert s_i \rangle \langle s_i \vert T^e \vert s'_j \rangle \langle s'_j \vert T^o\vert s_j \rangle \langle s_j \vert \cdots \vert s'_{n+1} \rangle \langle s'_{n+1} \vert R \rangle
\end{equation}
\end{widetext}
The matrix representations for the base vectors can be chosen as
\begin{equation}
\vert 1^+ \rangle=\vert 0^+ \rangle=\left(\begin{array}{c}
  1\\ 0
  \end{array}\right) \; , \;
\vert 1^- \rangle=\vert 0^- \rangle=\left(\begin{array}{c}
  0 \\ 1
  \end{array}\right).
\end{equation}
We assign these base vectors to the different vertices of each path according to the FIG. (\ref{fig2}).
Now according to the definition of the model only the following transitions are allowed
\begin{equation}
\begin{array}{lll}
\langle 1^+ \vert T^e \vert 0^+ \rangle=\frac{1}{z_1} &,& \langle 1^+ \vert T^e \vert 0^- \rangle=0 \\
\langle 1^- \vert T^e \vert 0^+ \rangle=0 &,& \langle 1^- \vert T^e \vert 0^- \rangle=\frac{1}{z_2} \\
\langle 0^+ \vert T^o \vert 1^+ \rangle=1 &,& \langle 0^+ \vert T^o \vert 1^- \rangle=1 \\
\langle 0^- \vert T^o \vert 1^+ \rangle=0 &,& \langle 0^- \vert T^o \vert 1^- \rangle=1 \\
\langle 0^+ \vert R \rangle =1 & , & \langle 0^- \vert R \rangle =1\\
\langle L \vert 0^+ \rangle =1 & , & \langle L \vert 0^- \rangle =0.
\end{array}
\end{equation}
These equations determine the transfer matrices $T^o$ and $T^e$
\begin{equation}
\label{TM}
T^o=\left(
\begin{array}{cc}
1 & 1 \\
0 & 1
\end{array} \right),\;
T^e=\left(
\begin{array}{cc}
\frac{1}{z_1} & 0 \\
0 & \frac{1}{z_2}
\end{array} \right)
\end{equation}
and also the vectors $\langle L \vert$ and $\vert R\rangle$
\begin{equation}
\label{Vectors}
\langle L \vert = (1, 0) \; , \;
\vert R \rangle= \left( \begin{array}{c}
1 \\ 1 \end{array} \right).
\end{equation}
Now using (\ref{TM}) and (\ref{Vectors}) one can readily see that (\ref{RPFTM}) gives (\ref{RPF}).\\
Following \cite{BGR04} we define two contact operators $\hat{a}_i=\vert 1^+ \rangle_{2i-1} \langle 0^+ \vert_{2i}$ and $\hat{b}_i=\vert 1^- \rangle_{2i-1} \langle 0^- \vert_{2i}$ associated with a contact from above and below the $x$-axis respectively. Now one can calculate the probability of finding a contact at site $2i$ above or below the horizontal line as follows
\begin{eqnarray}
\label{POCA}
\langle \hat{a}_i \rangle _n&=&\frac{\langle L \vert T^{i}\hat{a}_i T^{n-i} \vert R \rangle}{\langle L \vert T^n \vert R \rangle},\\
\label{POCB}
\langle \hat{b}_i \rangle _n&=&\frac{\langle L \vert T^{i}\hat{b}_i T^{n-i} \vert R \rangle}{\langle L \vert T^n \vert R \rangle}.
\end{eqnarray}
It is now easy to check that these probabilities can be written as
\begin{eqnarray}
\langle \hat{a}_i \rangle_n&=&\frac{\widetilde{Z}_{i}(z_1,\infty)\widetilde{Z}_{n-i}(z_1,z_2)}{\widetilde{Z}_{n}(z_1,z_2)},\\
\langle \hat{b}_i \rangle_n&=&\frac{\widetilde{Z}_{i-1}(z_1,z_2)\widetilde{Z}_{n-i+1}(\infty,z_2)}{\widetilde{Z}_{n}(z_1,z_2)}.
\end{eqnarray}
Using the fact that $\widetilde{Z}_n(z_1,z_2)=\widetilde{Z}_n(z_2,z_1)$, it can be seen that these probabilities have the following symmetry
\begin{equation}
\langle \hat{a}_{i} \rangle _n(z_1,z_2)=\langle \hat{b}_{n-i+1} \rangle _n(z_2,z_1)
\end{equation}
which can be understood from the symmetry of the paths. Let us define the average number of contacts from above and below the horizontal axis at position $x=\frac{i}{n}$ ($0\leq x \leq 1$) as
\begin{equation}
\label{ANC}
\langle \hat{a}_x \rangle = \langle \hat{a}_{xn} \rangle_n \;\;,
\;\;\langle \hat{b}_x \rangle = \langle \hat{b}_{xn} \rangle_n.
\end{equation}
Now in the thermodynamic limit $n \rightarrow \infty$ and in different phases of the one-transit walk model, we find for average number of contacts from above $x$-axis
\begin{equation}
\label{CFA}
\langle \hat{a}_x \rangle =
\left\{
\begin{array}{ll}
0 \;\; \mbox{for} \;\; z_1 > z_2,\\
1 \;\; \mbox{for} \;\; z_2 > z_1,\\
1-x \;\; \mbox{for} \;\; z_1 = z_2
\end{array}
\right.
\end{equation}
and for the average number of contacts from below $x$-axis
\begin{equation}
\label{CFB}
\langle \hat{b}_x \rangle =
\left\{
\begin{array}{ll}
1 \;\; \mbox{for} \;\; z_1 > z_2,\\
0 \;\; \mbox{for} \;\; z_2 > z_1,\\
x \;\; \mbox{for} \;\; z_1 = z_2.
\end{array}
\right.
\end{equation}
In the following sections we introduce a general one-dimensional driven-diffusive system with open boundaries. We discuss the case when a single product shock measure can develop in the system. The steady-state probability distribution will be given using a matrix method. Finally we investigate its connections with the walk model.
\section{Shock in a driven-diffusive system}
Consider a driven-diffusive system of classical particles defined on a one-dimensional lattice of length $n$ with open boundaries. For simplicity we consider the case where each lattice site can be occupied by a single particle or it can  be empty, although it is easy to generalize the idea to the systems with more than one species of particles. The time evolution of the probability distribution function of any configuration of a Markovian interacting particle system $\vert P(t)\rangle$ is governed by a master equation which can be written as a Schr\"odinger-like equation in imaginary time
\begin{equation}
\label{ME}
\frac{d}{dt}\vert P(t) \rangle =H \vert P(t) \rangle
\end{equation}
in which $H$ is called the Hamiltonian \cite{Schu00}. The matrix elements of the Hamiltonian $H$ are the transition rates between different configurations. For a system with nearest-neighbor interactions the Hamiltonian $H$ has the following general form
\begin{equation}
\label{Hamiltonian} H=\sum_{k=1}^{n-1}h_{k,k+1}+h_1+h_n.
\end{equation}
in which
\begin{eqnarray}
\label{Hamiltoniandet} h_{k,k+1}&=&{\cal I}^{\otimes (k-1)}\otimes h
\otimes {\cal I}^{\otimes (n-k-1)} \\ h_1 &=& h^{(l)} \otimes {\cal
I}^{\otimes (n-1)}
\\ h_n &=& {\cal I}^{\otimes (n-1)}\otimes h^{(r)}.
\end{eqnarray}
For our two-states system ${\cal I}$ is a $2 \times 2$ identity matrix and $h$ is a $4 \times 4$ matrix describing the bulk interactions. The two matrices $h^{(r)}$ and $h^{(l)}$ are both $2 \times 2$ matrices which describe the interactions at the boundaries.\\
A shock in such a system is defined as a sharp discontinuity between a hight-density and a low-density region. We assume that a single product shock measure in the system defined as
\begin{equation}
\label{PSM}
\vert k \rangle= \left( \begin{array}{c} 1-\rho_1 \\
\rho_1 \end{array} \right)^{\otimes k} \otimes \left( \begin{array}{c} 1-\rho_2 \\
\rho_2 \end{array} \right)^{\otimes n-k},
\end{equation}
in which $k \; (0\leq k\leq n)$ is the position of the shock, can evolve in time generated by the Hamiltonian of the system. The quantities $\rho_1$ and $\rho_2$ are the densities of the particles on the left-hand side and the right-hand side of the shock position. In \cite{JM07} the authors have studied the case where the shock position in (\ref{PSM}) has a simple random walk dynamics in the bulk of the system, while reflecting from the left and the right boundaries, by applying some constraints on the microscopic reaction rates of the system. If the shock front hops to the left (right) with the rate $\delta_l$ ($\delta_r$), we assume that it reflects from the right (left) boundary with the same rate. As we will discuss this will not alter the generality of the problem.\\
The steady-state of the system it this case can be written as a linear superposition of the shocks (\ref{PSM}). In the same reference the authors have also shown that the same steady-state can be obtained using a matrix-product approach. According to this approach the steady-state of the system $\vert P^*\rangle$ can be written as
\begin{equation}
\label{SS}
\vert P^* \rangle =\frac{1}{Z_n} \langle W \vert \left( \begin{array}{c} D \\
E \end{array} \right)^{\otimes n} \vert V \rangle
\end{equation}
provided that one chooses the operators $E$ and $D$, associated with the presence of a hole and a particle at each lattice site, as follows
\begin{equation}
E=\left( \begin{array}{cc}
(1-\rho_1) & d_0\\
0 & \frac{\delta_{l}}{\delta_{r}}(1-\rho_2)\\
\end{array} \right),
D=\left( \begin{array}{cc}
\rho_{1} & -d_0\\
0 & \frac{\delta_{l}}{\delta_{r}}\rho_{2}\\
\end{array} \right).
\end{equation}
The matrix elements of the vectors $\vert V \rangle$ and $\langle W \vert$, defined as
\begin{equation}
\label{Cond1}
\vert V \rangle =\left( \begin{array}{c}
v_1\\
v_2\\
\end{array} \right) \;,\; \langle W \vert=\left( \begin{array}{cc}
w_1&w_2\\
\end{array} \right)
\end{equation}
besides $d_0$ should only satisfy two constraints
\begin{equation}
d_0\frac{w_1}{w_2}=\rho_2-\rho_1 \;\;,\;\;d_0\frac{v_2}{v_1}=(\frac{\delta_l}{\delta_r})(\rho_1-\rho_2).
\end{equation}
The normalization factor in (\ref{SS}) is given by $Z_n=\langle W \vert (D+E)^n\vert V \rangle$. One should note that the only difference between the representation introduced here with the one used in \cite{JM07} is that here we have chosen the {\it upper triangular matrices} instead of {\it lower triangular matrices}.
\section{Connection between the walk model and the driven-diffusive system}
In what follows we show that the matrix-product partition function defined in the previous section is exactly equal to the partition function of the walk model $\widetilde{Z}_n(z_1,z_2)$ defined in (\ref{RPFTM}) provided that the fugacities of the contact point are chosen as
\begin{equation}
z_1=1 \;\;,\;\;z_2=\frac{\delta_r}{\delta_l}.
\end{equation}
Defining $C:=D+E$ and using a similarity transformation we can write
$$
Z_n=\langle \widetilde{W} \vert \widetilde{C}^n\vert \widetilde{V} \rangle
$$
in which
\begin{equation}
\widetilde{C}=U^{-1}CU\;,\;
\vert \widetilde{V} \rangle=U^{-1}\vert V \rangle\;,\;
\langle \widetilde{W} \vert=\langle W \vert U.
\end{equation}
We introduce $U$ as
\begin{equation}
U=\left( \begin{array}{cc}
u_1 & \frac{\frac{\delta_l}{\delta_r}}{1-\frac{\delta_l}{\delta_r}}u_1\\
0 & u_2\\
\end{array} \right)
\end{equation}
in which $u_1$ and $u_2$ are assumed to be non-zero. Since the matrix elements of the vectors $\langle W \vert$ and $\vert V \rangle$ should only satisfy (\ref{Cond1}) we define these elements as
\begin{equation}
v_1=\frac{u_1}{1-\frac{\delta_l}{\delta_r}}\;,\;v_2=u_2\;,\;w_1=\frac{1}{u_1}\;,\;
w_2=\frac{1}{u_2}(\frac{-\frac{\delta_l}{\delta_r}}{1-\frac{\delta_l}{\delta_r}})
\end{equation}
provided that
\begin{equation}
\label{Cond2}
d_0\frac{u_2}{u_1}=(\rho_1-\rho_2)(\frac{\frac{\delta_l}{\delta_r}}{1-\frac{\delta_l}{\delta_r}}).
\end{equation}
Now it is easy to verify that
\begin{equation}
\widetilde{C}=T\;,\;\vert \widetilde{V} \rangle=\vert R \rangle\;,\;
\langle \widetilde{W} \vert=\langle L \vert
\end{equation}
which means that the partition function of the walk model $\widetilde{Z}_n(z_1=1,z_2=\frac{\delta_r}{\delta_l})$ is equal to the normalization factor of the driven-diffusive system obtained from the matrix-product approach in (\ref{SS}).\\
Finally we show that the density profile of the particles in the driven-diffusive system in terms of the matrix-product approach defined as
\begin{equation}
\langle \rho_i \rangle=\frac{\langle W \vert C^{i-1}DC^{n-i}\vert V \rangle}{\langle W \vert C^{n}\vert V \rangle} \;, \; 1\leq i \leq n
\end{equation}
can be expressed in terms of the probability of finding a contact above or below the $x$-axis defined in (\ref{POCA}) and (\ref{POCB}). Using the same similarity transformation mentioned above one finds
\begin{equation}
\langle \rho_i \rangle =\frac{\langle L \vert T^{i-1} \widetilde{D} T^{n-i} \vert R \rangle }{\langle L \vert T^n \vert R \rangle}
\end{equation}
in which $\widetilde{D}=U^{-1}DU$. Now it is a straightforward step to show that the density profile of the particles can be rewritten as
\begin{equation}
\label{DP}
\langle \rho_i \rangle=\rho_1 \langle \hat{a_i} \rangle_n+\rho_2 \langle \hat{b_i} \rangle_n
\end{equation}
given that (\ref{Cond2}) is satisfied. Note that in terms of the fugacities $z_1$ and $z_2$ the phase diagram of both systems can be explained as follows: for $z_1 > z_2$ ($z_1 < z_2$) and in the thermodynamic limit $n \rightarrow \infty$ the walk has only contacts from below (above) the horizontal axis. This is equivalent to the fact that the shock front has a tendency to move to the left (right) which in turns means that the mean density of the particles in the bulk of the driven-diffusive system is equal to $\rho_2$ ($\rho_1$). One can see that these are confirmed by (\ref{DP}). On the other hand on the coexistence line one has $z_1=z_2$ or in terms of the shock position hopping rates $\delta_r=\delta_r$. On this line the shock position can be anywhere on the lattice with equal probabilities which results in a linear density profile for the particles on the lattice. Using (\ref{CFA}) and (\ref{CFB}) one can see that (\ref{DP}) gives the right density profile.
\section{Conclusion}
In this paper we considered two different systems: a one-transit walk model defined on a diagonally-rotated square lattice and a general driven-diffusive system in which a single product shock measure has a simple random walk dynamics. We showed that how the phase diagrams of these two systems are connected by associating the fugacities in the walk model with the hopping rates of the shock front in the driven-diffusive system. In fact, following the idea of \cite{BGR04} one can consider the normalization factor of a non-equilibrium system as a partition function of a two-dimensional equilibrium model. The fugacities defined in the walk model are related to the hopping rate of the shock front. The main difference of this paper with those previous is that our example belongs to a larger family of driven-diffusive systems instead of just the ASEP as a very special example. The physical quantities of these two systems are directly related (for instance see (\ref{DP})). As a simple generalization, it can be shown that by redefining the fugacities of the first and the last steps in the walk model, one can also generate the partition function of the driven-diffusive systems in the case where the hopping rates of the shock front in the bulk are different from those of the boundaries. As another generalization to this work one can try to investigates the driven-diffusive systems in a discrete time updating scheme and find the proper walk models which are related to them.



\begin{thebibliography}{1}

\bibitem{Schm95}B. Schmittmann and R.K.P. Zia, in: Phase Transitions and Critical
Phenomena Vol. 15, ed C. Domb and J.L. Lebowitz (Academic, London, 1995)

\bibitem{Priv97}V. Privman, non-equilibrium Statistical Mechanics in One
Dimension, (Cambridge University Press, Cambridge, 1997)

\bibitem{Schu00}G.M. Sch\"utz, in: Phase Transitions and Critical Phenomena
Vol. 19, ed C. Domb and J. L. Lebowitz (Academic, London, 2001)


\bibitem{BS02} V. Belitsky and G. M. Sch\"utz, {\it El. J. Prob.} {\bf 7} Paper No.11 1 (2002)

\bibitem{KJS03} K. Krebs, F. H. Jafarpour and G. M. Sch\"utz, {\it New Journal of Physics}
{\bf 5} 145.1-145.14 (2003)

\bibitem{PS04} M. Paessens and G.M. Sch\"utz, {\it New Journal of Physics} {\bf 6} 120 (2004)

\bibitem{RS04} A. R\'akos and G. M. Shc\"utz, {\it J. Stat. Phys.} {\bf 117} 55 (2004)

\bibitem{TS061} F. Tabatabaei, G.M. Sch\"utz, {\it Phys. Rev.} E {\bf 74} 051108 (2006)

\bibitem{TS062} F. Tabatabaei, G.M. Sch\"utz, {\it Diffusion Fundamentals} {\bf 4} 5.1-5.38 (2006)

\bibitem{JM09} F. H. Jafarpour and  S. R. Masharian, {\it Phys. Rev.} E {\bf 79} 051124 (2009)


\bibitem{BE07} R. A. Blythe and M. R. Evans, {\it J. Phys. A: Math. Theor.} {\bf 40} R333 (2007)

\bibitem{JM07} F. H. Jafarpour and  S. R. Masharian, {\it J. Stat. Mech.} P10013 (2007)


\bibitem{A00} P. F. Arndt, {\it Phys. Rev. Lett} {\bf 84} 814 (2000)

\bibitem{BE04} R. Brak and J. W. Essam, {\it J. Phys. A: Math. Gen.} {\bf 37} 4183 (2004)

\bibitem{BGR04} R. Brak, J. de Gier and V. Rittenburg, {\it J. Phys. A: Math. Gen.} {\bf 37} 4303 (2004)

\bibitem{BJJK04s} R.A. Blythe, W. Janke, D.A. Johnston and R. Kenna, {\it J. Stat. Mech.: Theory Exp.} P06001 (2004)

\bibitem{BJJK04p} R.A. Blythe, W. Janke, D.A. Johnston and R. Kenna, {\it J. Stat. Mech.: Theory Exp.} P10007 (2004)


\end{thebibliography}
\end{document}